\documentclass
[twocolumn,superscriptaddress,secnumarabic,amssymb,amsmath,nobibnotes,aps,prd,showpacs,nofootinbib]{revtex4}%
\usepackage{graphicx}
\usepackage{epsf}
\usepackage{bm}
\usepackage{amsmath}
\usepackage{amsfonts}
\usepackage{amssymb}
\usepackage{epstopdf}
\usepackage{natbib}
\usepackage{rotating}
\usepackage{pdflscape}
\usepackage{adjustbox}
\usepackage{color}
\usepackage{dcolumn}
\usepackage{ulem}
\usepackage{times}
\usepackage{appendix}
\usepackage{enumitem}
\usepackage{booktabs}
\usepackage{multirow}
\usepackage{orcidlink}

\begin{document}
\title{Bayesian deep learning for dark energy}

\author{Celia Escamilla-Rivera\orcidlink{0000-0002-8929-250X}}
\email{celia.escamilla@nucleares.unam.mx}
\affiliation{Instituto de Ciencias Nucleares, Universidad Nacional Aut\'onoma de M\'exico, Circuito Exterior C.U., A.P. 70-543, M\'exico D.F. 04510, M\'exico.}


%%%%%%%%%%%%%%%%%%%%%%%%%%%%%%%%%%%%%%%%%%%%
%%%%%%%%%%%%%%%%%%%%%%%%%%%%%%%%%%%%%%%%%%%%
\begin{abstract}
In this chapter we discuss basic ideas on how to structure and study the Bayesian methods for standard models of dark energy and how to implement them in the architecture of deep learning processes.  
\end{abstract}

\maketitle

%%%%%%%%%%%%%%%%%%%%%%%%%%%%%%%%%%%%%%%%%%%%
%%%%%%%%%%%%%%%%%%%%%%%%%%%%%%%%%%%%%%%%%%%%

\section{Introduction}
\label{sec:intro}

The dark sector of our universe has been the problem of study for cosmologists who are striving to found an answer to both dark matter and dark energy. While we do have estimates of the likely percentages of baryonic matter, dark matter, and dark energy at 5\%, 27\%, and 68\%, respectively, we have been trying to improve these quantities and optimise the numerical expense of the statistical methods employed to analyse cosmological data currently available.

These thoughts have opened the path of the following chapter, where we will discuss from the standard dark energy models to explain the cosmic acceleration until the design of a numerical architecture to understand the constraints over the cosmological parameters that can describe the current universe and its effects. 

A highlight in the observed universe is the origin and nature of the
cosmic accelerated expansion.
The standard cosmological model that is consistent with current cosmological observations is the standard concordance model or $\Lambda$CDM.
According to this model, the observed current cosmic acceleration is related to the
repulsive gravitational force of a Cosmological Constant $\Lambda$ with constant energy density $\rho$ and negative pressure
$p$. This proposal has been the backbone of the standard cosmology since the nineties, but simple enough as it is
the proposal has a couple of theoretical problems, two of them are  the fine-tuning argument and
coincidence problem \cite{Weinberg:2000yb,Sahni:1999gb}.
To found a solution to these problems, some ideas have lead to new
theories that can modify the General
Relativity  (GR) or consider a scenario with a dynamical dark energy fluid. 
It is in this way that dark energy emerges as a solution since it can be described as an exotic fluid 
parametrised by an equation of state (EoS), which can be written in terms of the redshift, $w(z)$. 
Until now, the properties of this EoS remains
under-researched. There have been several proposals on dark energy parameterisations
discussed in the literature (see, e.g., \cite{Feng:2011zzo,Stefancic:2005sp,Jassal:2004ej,Wang:2008zh,Wang:2005yaa}), addressing from parameterisation as Taylor-like polynomials to dynamical EoS  that can provide oscillatory behaviours.

Nowadays, the techniques to discriminate between models and confront them with $\Lambda$CDM, are based on the 
calculations of the constraints on the EoS free parameter(s) of the models. This methodology has been done using observables 
 that can show the cosmic acceleration such as supernovae type IA (SNeIa), baryon acoustic
oscillations (BAO), cosmic microwave background (CMB), weak lensing spectrum, etcetera. 
Astrophysical measurements, e.g Pantheon from supernovae \cite{Scolnic:2017caz}, or BAO from BOSS \cite{Busca:2012bu}, allow us to constrain the cosmological parameters for a specific model. Also, with them we can test deviations from the standard $\Lambda$CDM model. Along these years, there have been many surveys related to the test of the cosmic acceleration, e.g from Union2.1 \footnote{\url{http://supernova.lbl.gov/Union/}} to the Joint LightCurve Analysis \cite{Ade:2015xua,Betoule:2014frx}. Moreover, the statistics have been improved due to the density of data these kinds of supernovae.

%%%%%%%%%%%%%%%%%%%%%%%%%%%%%%%%%%%%%%%%%%%%%
%%%%%%%%%%%%%%%%%%%%%%%%%%%%%%%%%%%%%%%%%%%%%

\section{On how to model dark energy}
\label{sec:Mod_DE}

One of the first steps to understand the behaviour of the cosmic acceleration remains in that we require an energy density
with negative pressure at late times. To achived this we need to express the ratio between the pressure and energy density as negative, i.e., $w(z)= P/\rho <0$. In order to develop the evolution equations for a universe with this kind of \textit{fluid}, we start by introducing in Einstein equations a Friedmann-Lemaitre-Robertson-Walker metric to obtain the
Friedmann and Raychaudhuri equations for a spatially
flat universe
\begin{eqnarray} 
 E(z)^2=\left(\frac{H(z)}{H_0}\right)^2
 &=& \frac{8\pi G}{3}(\rho_m + \rho_{DE})  \Big[\Omega_{0m} (1+z)^3 
 \nonumber\\&&
 +\Omega_{0(DE)} f(z)\Big], \label{hubble eq}
 \end{eqnarray}
and
\begin{equation}
 \frac{\ddot{a}}{a} =
 -\frac{H^2}{2}\left[\Omega_m +\Omega_{DE} (1 + 3w)\right],\label{hubble eq2}
\end{equation}
where $H(z)$ is the Hubble parameter in terms of the redshift $z$, $G$ the gravitational constant and the subindex $0$ indicates the present-day values for the Hubble parameter and matter densities.

From (\ref{hubble eq2}) it is possible to obtain the energy conservation equation, in that way
the energy density of the non-relativistic matter is $\rho_m (z)=\rho_{0m} (1+z)^3$. And the
$\rho_m$ is given by:
\begin{equation}
 \begin{aligned}
  \rho_m (z)=\rho_{0m} (1+z)^3,
 \end{aligned}
\end{equation}
and the
dark energy density can be modelated as $\rho_{DE} (z)= \rho_{0(DE)} f(z)$, where
can be written as:
\begin{equation}
 \begin{aligned}
  \rho_{DE} (z)= \rho_{0(DE)} f(z).
 \end{aligned}
\end{equation}
If we assuming that the energy-momentum tensor (on the right side of the Einstein's equations) $T^{\mu\nu}$ is a perfect fluid (without viscosity or stress effects), i.e. $\nabla_\mu T^{\mu\nu}=0$, the form of $f(z)$ can be restricted to be:
\begin{equation}
f(z)= e^{\left[3\int^{z}_{0}{\frac{1+w(\tilde{z})}{1+\tilde{z}}}d\tilde{z}\right]}. \label{eq:eos_g}
\end{equation}

Now, the behaviour of the latter is restricted directly to the form of 
$w(z)$, which can give a description
of the Hubble function (which can be normalised by the constant Hubble $H_0$), as e.g., in the case of
quiessence models ($w = const.$) the solution of $f(z)$ is
$f(z)=(1+z)^{3(1+w)}$. If we consider the case of the cosmological constant ($w=-1$) then $f =1$. 

Some interesting insights of the above forms for $w(z)$ has been reported in  \cite{Stefancic:2005sp,Lazkoz:2005sp} and references therein, where
a dark energy density $\rho_{DE}$ with
varying and non-varying $w(z)$ are considered.

As an extension, with the later equations we can calculate the dynamical age of the universe using the follow relationship:
\begin{equation}
\Omega_m +\Omega_{DE} =1 \quad \text{or} \quad \frac{\rho_m}{\rho_{DE}}=\frac{\Omega_m}{\Omega_{DE}}.
\end{equation}
Integrating we can obtain
\begin{equation}
 t_0 = \int^{\infty}_{0}{\frac{dz}{(1+z)H(z)}},
\end{equation}
\begin{equation} 
  t_0 = {H_0}^{-1}\int^{\infty}_{0}{\frac{dz}{(1+z)\sqrt{\left[\Omega_{0m} (1+z)^3 +\Omega_{0(DE)} f(z)\right]}}}. \label{age_universe}
\end{equation}
From here we can set
a  functional form of $f(z)$, which contribution
of the dark energy density to $H(z)$ in (\ref{hubble eq}) goes to a region of
negative values of $w(z)$. The physics behind this behaviour
is an impact on the evolution
of dark energy using the dynamical age of the universe (\ref{age_universe}).
 When we compare several theoretical models in the light of observations a model approach is essential. As we mentioned in the Introduction,
 to obtain a dark energy model with late-time negative pressure
we can think in two scenarios:
\begin{itemize}
 \item a quiessence model: which can show a wide
application in tracker the slow roll condition of scalar fields and
demands a constant value of $w$. As an example, for a flat universe and according to the
Planck data \cite{Ade:2015xua}, the dark energy EoS parameter gives $w=-1.006\pm 0.045$, which is
consistent with the cosmological constant. This data constrain the curvature parameter at 2$\sigma$
and is found to be very close to zero with $|\Omega_k|< 0.005$.
\item a kinessence model; where
when the EoS is a function of redshift $z$. For this case, several dark energy models
with different parameterisations of $w(z)$ has been discussed
in the literature \cite{Lazkoz:2005sp}. 
\end{itemize}

%%%%%%%%%%%%%%%%%%%%%%%%%%%%%%%%%%%%%%%%%%%%%
%%%%%%%%%%%%%%%%%%%%%%%%%%%%%%%%%%%%%%%%%%%%%
\section{Standard dark energy Models}
\label{sec:bidim-DE}

One of the most commonly used proposals in the literature are 
Taylor series-like parameterisations:
\begin{equation}
w(z) = \sum_{n = 0}{w_nx_n(z)},
\end{equation}
where $w_n$ are constants and $x_n(z)$ are functions of the redshift $z$, or, the scalar factor $a$. As brief examples, in this section we present three models that have bidimensional forms in the since that they depend only of two free parameters $w_i$. A first target is to express the exactly form of the Hubble function using a specific expression for $w$ given by (\ref{eq:eos_g}). Once integrated, we can normalise this function by a Hubble parameter $H_0$, from now on we called this normalisation function depending of the redshift as $E(z)=H(z)/H_0$. The second target is to test these equations with 
the current astrophysical data available.

%%%%%%%%%%%%%%%%%%%%%%%%%%%%%%%%%%%%%%%%%%%%%
%%%%%%%%%%%%%%%%%%%%%%%%%%%%%%%%%%%%%%%%%%%%%
\subsection{Lambda Cold Dark Matter-Redshift parameterisation ($\Lambda$CDM)}

This standard cosmological model is represented by:
\begin{equation}
 E(z)^2=\Omega_m (1+z)^3 + (1-\Omega_m), \label{LCDM}
\end{equation}
where $\Omega_m$ is the matter density (including the non-relativistic and dark matter components). For this model, the value of $w=-1$.
As it is well known in the literature, this model provides a good fit for several number of observational astrophysical data
without experimenting with the theoretical problems mentioned in the Introduction. 

%%%%%%%%%%%%%%%%%%%%%%%%%%%%%%%%%%%%%%%%%%%%%
%%%%%%%%%%%%%%%%%%%%%%%%%%%%%%%%%%%%%%%%%%%%%
\subsection{Linear-Redshift parameterisation (LR)}

One of the first attempts using Taylor series --at first order-- is the EoS given by \cite{Huterer:2000mj,Weller:2001gf} 
\begin{equation}\label{linear}
\begin{aligned}
w(z) = w_0 - w_1 z,
\end{aligned}
\end{equation}
from we can recover $\Lambda CDM$ model if $w(z)=w=-1$ with $w_0 =-1$ and $w_1 =0$. We notice immediately that
due the linear term in $z$, this proposal diverges at high redshift and consequently yields strong constraints on $w_1$ in
studies involving data at high redshifts, e.g., when we use CMB data \cite{Wang:2007dg}.

As usual, we can use the later to obtain an expression for the Hubble normalised function as:
\begin{eqnarray}\label{linear p}
 E(z)^2 &=& \Omega_m (1+z)^3
 +(1-\Omega_m)(1+z)^{3(1+w_0 +w_1)}
\nonumber\\&&
\times e^{-3w_1 z}
\end{eqnarray}

%%%%%%%%%%%%%%%%%%%%%%%%%%%%%%%%%%%%%%%%%%%%%
%%%%%%%%%%%%%%%%%%%%%%%%%%%%%%%%%%%%%%%%%%%%%
\subsection{Chevallier-Polarski-Linder Parameterization (CPL)}

Due the consequence of the LP parameterisation divergence, Chevallier, Polarski and Linder proposed a
simple parameterisation \cite{Chevallier:2000qy,Linder:2007wa} that in particular can be
represented by two $w_i$ parameters that are given by a present value of the EoS $w_0$ and its overall time evolution $w_1$. The
proposal is given by the expression
\begin{eqnarray}
 w (z)= w_0 +\left(\frac{z}{1+z}\right) w_1,  \label{eq:cpl}
\end{eqnarray}
and its evolution is
\begin{eqnarray}\label{CPL}
 E(z)^2&=& \Omega_m (1+z)^{3}  +(1-\Omega_m)(1+z)^{3(1+w_0 +w_1)}
 \nonumber\\ &&
 \times e^{-\left(\frac{3w_1 z}{1+z}\right)}.
\end{eqnarray}
As we can notice, the divergence at high redshift relax, but still this ansatz have some problems in specific low redshift range of observations.

%%%%%%%%%%%%%%%%%%%%%%%%%%%%%%%%%%%%%%%%%%%%%
%%%%%%%%%%%%%%%%%%%%%%%%%%%%%%%%%%%%%%%%%%%%%

\section{Estimating the cosmological parameters}

After we have defined a specific cosmological model, we can then perform their test using astrophysical observations. The methodology can be described
by a simple calculation of the usual $\chi^2$ method and then process the MCMC chains computational runs around a certain value (observational(s) point(s)), and 
obtain the best fit parameter(s) of this process.  Parameter estimation is usually done by computing the so-called \textit{likelihood function} for several values of the cosmological parameters. For each data points in the parameter space, the likelihood $\mathcal{L}$ function gives the minimise probability of obtaining the observational data that was obtained if the hypothesis parameters had the given values (or priors). For example, the standard cosmological model $\Lambda$CDM is described by six parameters, which include the amount of dark matter and dark energy in the universe as well as its expansion rate $H$. Using the CMB data (which is the accuracy data that we understand very well so far), a likelihood function can be constructed. The information given by $\mathcal{L}$  can tell which values of these parameters are more likely, i.e by probing many different values. Therefore, we can to determine the values of the parameters and their uncertainties via error propagation over the free parameters of the model.

Now, the following question is what kind of astrophysical surveys\footnote{This word in the colloquial language also can be replaced by \textit{likelihood} --do not misunderstand with function  $\mathcal{L}$. Or simple we can called as \textit{samplers}. } can we use to test the cosmological models? In the next sections we described the most used surveys that are employed to analise the cosmic acceleration. It is important to mention that these surveys spread depending on its nature. We have three types of observations classified as standard candels (e.g supernovae, which characteristic function is the luminosity distance), standard rulers (e.g supernovae, which characteristic function is the angular/volumen distance) and the standard sirens (e.g gravitational waves, which can be described by frequencies or chirp masses depending on the observation). The set of all of them can describe precise statistics, but by separate, each of them have intrinsic problems due to their physical definition. For supernovae, the luminosity distance has in their definition an integral of the cosmological model, therefore when we perform the error propagation, the uncertainty is high. This disadvantage can be compensated by the large population of data points in the sampler. On the other hand, the uncertainty is less for standard rulers in comparison to supernovae. In this case, the definition of angular distance does not include integrals. The price that we pay to use this kind of sampler is that the population of data is very small (e.g from surveys like BOSS or CMASS, we have only 7 data points). Moving forward,  the observation of gravitational wave standard sirens would be developed into a powerful new cosmological test due that they can play an important role in breaking parameter degeneracies formed by other observations as the ones mentioned. Therefore, gravitational wave standard sirens are of great importance for the future accurate measurement of cosmological parameters. In this part of the chapter, we are going only to develop the use of the first two kinds of observations.

%%%%%%%%%%%%%%%%%%%%%%%%%%%%%%%%%%%%%%%%%%%%%
%%%%%%%%%%%%%%%%%%%%%%%%%%%%%%%%%%%%%%%%%%%%%
\section{Supernovae sampler}
\label{sec:observations}

Over the ninety years, since their discovery, Type Ia supernovae (SNIa) have been the proof of the current cosmic acceleration. The surveys have been changing given us a large population of observations, from Union 2.1\footnote{\url{http://supernova.lbl.gov/Union/}} to the Joint LightCurve Analysis \cite{Ade:2015xua,Betoule:2014frx}, the data sets have been incrementing observations and also their redshift range. Currently, the Pantheon sampler, which consists of a total 1048 Type Ia supernovae (SNIa) in 40 bins \cite{Scolnic:2017caz} compressed, is the
largest spectroscopically confirmed SNIa sample to date. This latter characteristic makes this sample attractive to constrain with considerably precision the free cosmological parameters of a specific model.

SNIa can give determinations of the distance modulus $\mu$, whose theoretical prediction is related to the luminosity distance $d_L$ according to:
\begin{equation}\label{eq:lum}
\mu(z)= 5\log{\left[\frac{d_L (z)}{1 \text{Mpc}}\right]} +25,
\end{equation}
where the luminosity distance is given in units of Mpc. In the standard statistical analysis, one adds to the distance modulus the nuisance parameter $ \mu_0$, an unknown offset sum of the supernovae absolute magnitude (and other possible systematics), which is degenerate with $H_0$. 

Now, the statistical analysis of the this sample rests on the definition of the modulus distance as:
\begin{equation}
\mu(z_{j}, \mu_0) = 5 \log_{10} [ d_{L}(z_{j}, \Omega_m; \boldsymbol{\theta}) ] + \mu_0,
\end{equation}
where $d_{L}(z_{j}, \Omega_m; \boldsymbol{\theta})$ is the Hubble free luminosity
distance:
\begin{equation}\label{eq:dl_H}
d_{L}(z, \Omega_m; \boldsymbol{\theta}) = (1+z) \ \int_{0}^{z} \mathrm{d}z'
\frac{1}{E(z', \Omega_m; \boldsymbol{\theta})} \; .
\end{equation}
With this notation we expose the different roles of the several cosmological
parameters appearing in the equations: the matter density parameter $\Omega_m$ appears separated as it is assumed to be fixed
to a prior value, while $\boldsymbol{\theta}$ is the EoS parameters $w_i$. These later are the parameters that we will be constraining by the data. The best fits will be obtained by minimising the
quantity
\begin{equation}\label{eq: sn_chi}
\chi^{2}_{\mathrm{SN}}(\mu_{0}, \boldsymbol{\theta}) = \sum^{\mathcal{N}_{\mathrm{SN}}}_{j =
1} \frac{(\mu(z_{j}, \Omega_m; \mu_{0}, \boldsymbol{\theta})\} -
\mu_{obs}(z_{j}))^{2}}{\sigma^{2}_{\mathrm{\mu},j}},
\end{equation}
where the $\sigma^{2}_{\mathrm{\mu},j}$ are the measurement
variances.
And nuisance parameter $\mu_{0}$ encodes the Hubble
parameter and the absolute magnitude $M$, and has to be
marginalised over. 

We assume spatial flatness, where the luminosity distance is related to the comoving distance $D$ as
\begin{equation}
d_{L} (z) =\frac{c}{H_0} (1+z)D(z),
\end{equation}
where $c$ is the speed of light, so that, using (\ref{eq:lum}) we can obtain
\begin{equation}
D(z) =\frac{H_0}{c}(1+z)^{-1}10^{\frac{\mu(z)}{5}-5}.
\end{equation}
The function $E(z)$ can be calculated by considering $D(z)=\int^{z}_{0} H_0 d\tilde{z}/H(\tilde{z})$. Instead of using the entire set of parameters for the sampler, we can employ the Pantheon binned list for CosmoMC to constrain the models (analogous to the Joint Light Curve Analysis sampler \cite{Betoule:2014frx}).

Here, $M$ is the nuisance parameter n the sample, and we select respective values of $ \mu_0$ from a statistical analysis of the $\Lambda$CDM model with Pantheon observation obtained by fitting $H_0$ to the Planck value given in \cite{Aghanim:2018eyx}. This kind of fit using computational tools that can run standard MCMC chains. In cosmology --at least at the moment this text is writing-- several codes have been implemented to perform the statistical fit of this parameter. The lector can explore the tool called MontePython code \footnote{https://monte-python.readthedocs.io/en/latest/} and run a standard MCMC for $M$ using the model of their preference. As an example, if we run a $\Lambda$CDM model with this supernovae sample, the mean value obtained will be $ \mu_0=-19.63$.

%%%%%%%%%%%%%%%%%%%%%%%%%%%%%%%%%%%%%%%%%%%%%
%%%%%%%%%%%%%%%%%%%%%%%%%%%%%%%%%%%%%%%%%%%%%
\section{Baryon Acoustic Oscillation sampler}

As a standard rulers, these astrophysical observations can contribute important features by comparing the data of the
sound horizon today to the sound horizon at the time of recombination (extracted from the CMB anisotropy~data).
Usually, the baryon acoustic distances are given as a combination of the angular scale and the redshift separation. 

To define these quantities we 
require a relationship via the ratio:
\begin{equation}
d_{z} \equiv \frac{r_{s}(z_{d})}{D_{V}(z)}, \quad \text{with} \quad  r_{s}(z_{d}) = \frac{c}{H_{0}} \int_{z_{d}}^{\infty}
\frac{c_{s}(z)}{E(z)} \mathrm{d}z
\end{equation}
where $r_{s}(z_{d})$ is the comoving sound horizon at the baryon dragging epoch,
\begin{equation}
r_{s}(z_{d}) = \frac{c}{H_{0}} \int_{z_{d}}^{\infty}
\frac{c_{s}(z)}{E(z)} \mathrm{d}z\; ,
\end{equation}
and $z_{d}$ is the
drag epoch redshift with $c^{2}_{s}= c^2/3[1+(3\Omega_{b0}/4\Omega_{\gamma 0})(1+z)^{-1}]$ as the sound speed
with $\Omega_{b0}$ and $\Omega_{\gamma 0}$, which are the present values of baryon and photon parameters, respectively. 

We define the dilation scale as
\begin{equation}
D_{V}(z,\Omega_m; w_0,w_1) = \left[ (1+z)^2 D_{A}^2 \frac{c
\, z}{H(z, \Omega_m; w_0,w_1)} \right]^{1/3},
\end{equation}
where $D_{A}$ is the angular diameter distance given by
\begin{equation}
D_{A}(z,\Omega_m; w_0,w_1) = \frac{1}{1+z} \int_{0}^{z}
\frac{c \, \mathrm{d}\tilde{z}}{H(\tilde{z}, \Omega_m;w_0,w_1)}. \;
\end{equation}

Using the comoving sound horizon, we can relate the distance ratio $d_{z}$ 
with the expansion parameter $h$ (defined such that {$H \doteq 100 h$)} and the physical densities $\Omega_{m}$ and $\Omega_{b}$.
Therefore, we have
\begin{equation}
r_{s}(z_{d}) = 153.5 \left( \frac{\Omega_{b}
h^2}{0.02273}\right)^{-0.134} \left( \frac{\Omega_{m}
h^2}{0.1326}\right)^{-0.255} \; \mathrm{Mpc}, \;
\end{equation}
with $\Omega_m= 0.295\pm0.304$ and $\Omega_b=0.045\pm0.00054$ \cite{Betoule:2014frx}. As we mentioned above, unfortunately so far we have a very low data population of this sampler. Moreover, as an example for this text, we 
employed compilations of three current surveys: $d_{z}(z=0.106)=0.336\pm 0.015$ from 6-degree Field Galaxy Survey (6dFGS) \cite{Beutler:2011hx},
$d_{z}(z=0.35)=0.1126\pm 0.0022$ from Sloan Digital Sky Survey (SDSS) \cite{Anderson:2013zyy} and $d_{z}(z=0.57)=0.0726\pm 0.0007$ from Baryon Oscillation Spectroscopic Survey (BOSS) with  high-redshift  CMASS \cite{Xu:2012hg}. 

We can also, add to the full sample three correlated measurements of $d_{z}(z=0.44)=0.073$, $d_z(z=0.6)=0.0726$ and $d_z(z=0.73)=0.0592$
from the WiggleZ survey \cite{Blake:2012pj}, which has the inverse covariance matrix:
\begin{equation}
\mathbf{C^{-1}_{WiggleZ}}=\left(\begin{array}{ccc}
1040.3 & -807.5 & 336.8\\
-807.5 & 3720.3 & -1551.9 \\
336.8 & -1551.9 & 2914.9\\
\end{array} \right)\;
\end{equation}

In order to perform the $\chi^2$-statistic, we define the proper $\chi^2$ function for the BAO data as
\begin{equation}\label{chibao}
\chi^2_{\mathrm{BAO}}(\boldsymbol{\theta}) =
\mathbf{X}^T_{\mathbf{BAO}}
\mathbf{C}^{-1}_{\mathbf{BAO}}
\mathbf{X}_{\mathbf{BAO}}
\end{equation}
where $\mathbf{X}_{\mathbf{BAO}}$ is given as
\begin{equation}
\mathbf{X_{BAO}}=\left(\begin{array}{c}
\frac{r_s (z_d)}{D_V (z,\Omega_m;w_0,w_1)})    - d_{z}(z)\\
\end{array} \right)\;
\end{equation}

Then, the total $\chi^2_{\mathrm{BAO}}$ is directly obtained by the sum of the individual quantity by using (\ref{chibao}) in
\begin{equation}
\chi^2_{\mathrm{BAO-total}}=\chi^2_{\mathrm{6dFGS}} +\chi^2_{\mathrm{SDSS}} +\chi^2_{\mathrm{BOSS CMASS}} +\chi^2_{\mathrm{WiggleZ}.}
\end{equation}

%%%%%%%%%%%%%%%%%%%%%%%%%%%%%%%%%%%%%%%%%%%%%
%%%%%%%%%%%%%%%%%%%%%%%%%%%%%%%%%%%%%%%%%%%%%
\section{How to deal with Bayesian statistics }
\label{sec:bayesian}

Now we are ready to introduce how to extrapolate the above frequentist analyses to the bayesian field. The important difference between both statistics is that in the first one we are dedicated in work with a standard $\chi^2$ fit, while in the second one we are taking into account the following idea: given a specific set of cosmological values (the priors), which are the probability of a second set of values to fit the hypothesis. 

The above idea is what we call a Bayesian model selection, which methodology consist in describe the relationship between the cosmological model,
the astrophysical data, and the prior information about the free parameters. Using Bayes theorem \cite{bayes-th}
we can update the prior model probability to the posterior model probability. However, when we compare
models, the evidence function is used to evaluate the model's evolution using the data at hand. 

We define the evidence function as:
\begin{equation}\label{eq:bayes}
\mathcal{E} =\int{\mathcal{L}(\theta) P(\theta) d\theta},
\end{equation}
where $\theta$ is the vector of free parameters (which for the dark energy models presented in the above sections, will be given by the $w_i$ free parameters). $P(\theta)$
is the prior distribution of these parameters. 

From a computational point of view, and due to the large population of data and the model used, (\ref{eq:bayes}) can be difficult to calculate since
the integrations can consume much computational time when the parametric phase space is large.
Nevertheless, even when several methods exist \cite{gregory,Trotta:2005ar}, in this text we present a test with a nested
sampling algorithm \cite{skilling} which has proven practicable in cosmology applications \cite{Liddle:2006kn}.

Once we obtain the evidence, we can, therefore, calculate the logarithm of the Bayes factor between two models $\mathcal{B}_{ij}=\mathcal{E}_{i}/\mathcal{E}_{j}$,
where the reference model ($\mathcal{E}_{i}$) with the highest evidence can be the $\Lambda$CDM model and impose a flat prior on $H_0$, i.e we can use an exact value of this parameter.

The interpretation of the results of this ratio can be described by a scale known as Jeffreys's scale \cite{jeffreys}, which easily can be explained as follow: 
\begin{itemize}
\item if
$\ln{B_{ij}}<1$ there is not significant preference for the model with the highest evidence;
\item  if $1<\ln{B_{ij}}<2.5$ the
preference is substantial;
\item and, if $2.5<\ln{B_{ij}}<5$ it is strong; if $\ln{B_{ij}}>5$ it is decisive.
\end{itemize}

%%%%%%%%%%%%%%%%%%%%%%%%%%%%%%%%%%%%%%%%%%%%%
%%%%%%%%%%%%%%%%%%%%%%%%%%%%%%%%%%%%%%%%%%%%%
\section{About deep learning in cosmology} 
\label{sec:DL}

Bayesian evidence method remains the preferred method compared with information criteria and Gaussian processes in the literature. A full Bayesian inference for model selection --in the case we have a landscape in where we can discriminate a pivot model from a hypothesis-- is computationally expensive and often suffers from multi-modal posteriors and parameter degeneracies. This latter issue leads to a large time consumption to obtain the final best fit for the free parameters.
As the study of the Large Scale Structure (LSS) of the universe indicates, all our knowledge relies on state-of-the-art cosmological simulations to address several questions by constraining the cosmological parameters at hand using Bayesian techniques. Moreover, due to the computational complexity of these simulations, some studies look to remain computationally infeasible for the foreseeable future. It is at this point where computational techniques as machine learning can have some important uses, even for trying to understand our universe.

The idea behind the machine learning is based in consider a neural network with a complex combination of neurons organised in nested layers. Each of these neuron implements a function that is parametrised by a set of weights $W$. And every layer of a neural network thus transforms one input vector --or tensor depending the dimension-- to another through a differentiable function. Theoretically, given a neuron $n$ it will receive an input vector and the choice of an activation function $A_n$, the output of the neuron can be computed as 
\begin{equation}
    h^{<t>}=A_n(h^{<t-1>}\cdot W_{h}+x^{<t>}\cdot W_{x}+b_{a}),
    \end{equation}
    \begin{equation}
    y^{t}=A_n(h^{t}\cdot W_{y}+b_{y}),
\end{equation}
where $ h^{<t>}$ is called the hidden state, $A_n$ is the activation function and $y^{t}$ is the output.

The goal to introduce a set of data to train this array and therefore the architecture can learn to finally give an output set of data. For example:
the network can learn the distribution of the distance moduli in the dark energy models, then feed the astrophysical samplers (surveys) to the network to reconstruct the dark energy model and then discriminate the most probable model. \footnote{In this text we are employing a Recurrent Neural Network. There are several in this machine learning field e.g. in \cite{Ntampaka:2019udw} and references therein. }  

Moreover, while neural networks can learn complex nested representations of the data, allowing them to achieve impressive performance results, it also limits our understanding of the model learned by the network itself. The choice of an architecture \cite{Ntampaka:2019udw} can have an important influence on the performance of the neural network. Some designs have to make concerning the number and the type of layers, as well as the number and the size of the filters used in each layer. A convenient
way to select these choices is typically through experimentation --which for our universe, we will need these to happen first.-- As it is, we can select the size of the network, which depends on the number of training tests as networks with a large number of cosmological parameters are likely to overfit if not enough training tests are available.

At the moment these lines are writing, a strong interest over this kind of algorithm is bringing new opportunities for data-driven cosmological discovery, but will also present new challenges for adopting machine learning --or, in our case, a subset of this field, deep learning-- methodologies and understanding the results when the data are too complex for traditional model development and fitting with statistics. A few proposals in this area has been done to explore the deep learning methods for measurements of cosmological parameters from density fields \cite{Schmelzle:2017vwd} and for future large-scale photometric surveys \cite{Charnock:2016ifh}.

%%%%%%%%%%%%%%%%%%%%%%%%%%%%%%%%%%%%%%%%%%%%%
%%%%%%%%%%%%%%%%%%%%%%%%%%%%%%%%%%%%%%%%%%%%%
\section{Deep learning for dark energy}
 
A first target to start to train an astrophysical survey is to design an architecture with an objective function of neural networks that can have many unstable points and a local minima. This architecture makes the optimisation process very difficult, but in real scenarios, high levels of noise degrade the training data and typically result in optimisation scenarios with more local minima and therefore increases the difficulty in training the neural network. It can thus be desirable to start optimising the neural network using noise-free data which typically yields smoother scenarios. As an example, in Figure \ref{fig:dl_de} we present a standard network using an image of a cosmological simulation (the data) and then divided an array of several layers to finally extract the output cosmological parameters value. Each neuron uses a Bayesian process to compute the error propagation as it is done in the standard inference analyses.

We can describe a quick, but effective, recipe to develop a Recurrent Neural Network with a Bayesian computation training in the following steps:
\begin{itemize}
        \item Step 1. Construction of the neural network. For a Recurrent Neural Network method we can choose values that have one layer and a certain number of neurons (e.g you can start with 100 for a supernovae sampler). 
          \item Step 2. Organising the data. We need to sort the sampler from lower to higher redshift in the observations. Afterward, we re-arrange our data using the number of steps (e.g try with 4 steps numbered as $x_{i}$ for a supernovae sampler). 
        \item Step 3. Computing the Bayesian training. Due to the easiness of neural networks to overfit, it is important to choose a mode of regularisation. With a Bayesian standard method to compute the evidence, the algorithm can calculate errors via regularisation methods\cite{Aurelien}. Finally, over the cost function we can use Adam optimiser.
        \item Step 4. Training the entire architecture. It is suitable to consider a high number of epochs (e.g for a sampler as Pantheon, you can try with 1000 epoch per layer). After the training, it is necessary to read the model and apply more times the same dropout to the initial model. The result of this step is the construction of the  confidence regions.
        \item Step 5. Computing modulus distance $\mu(z)$ for each cosmological model. Using the definitions of $E(z)$, we can compute $\mu(z)$ by using a specific dark energy equation of state in terms of $z$ and then integrating them. 
                \item Step 6. Computing the best fits. Finally, the output values can be obtained by using the training data as a simulated sample. We use the publicly codes CLASS \footnote{\url{https://github.com/lesgourg/class_public}} and Monte Python \footnote{\url{https://github.com/baudren/montepython_public}} to constrain the models as it is standard for usual Bayesian cosmology.
        \end{itemize}

\begin{widetext}
\begin{center}
\begin{figure*}
    \includegraphics[width=0.8\textwidth,origin=c,angle=0]{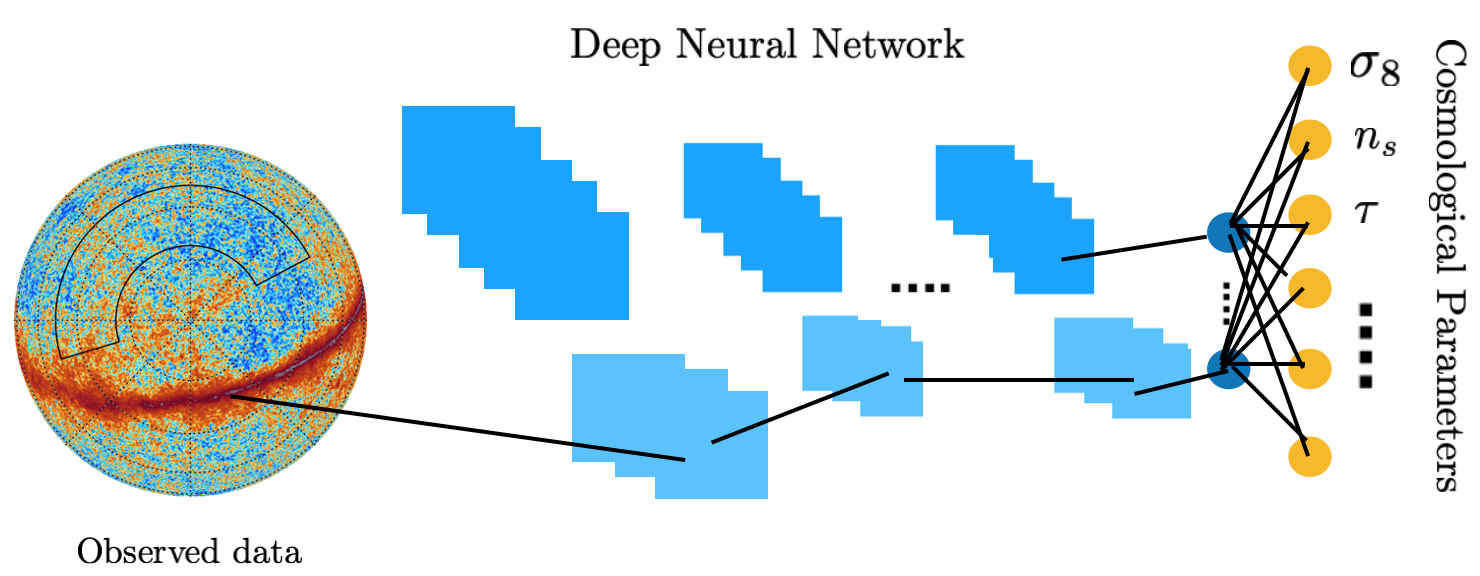}
    \caption{A deep learning architecture for dark energy.}
    \label{fig:dl_de}
\end{figure*}
    \end{center}
\end{widetext}

\begin{figure}
\centering
    \includegraphics[width=0.51\textwidth,origin=c,angle=0]{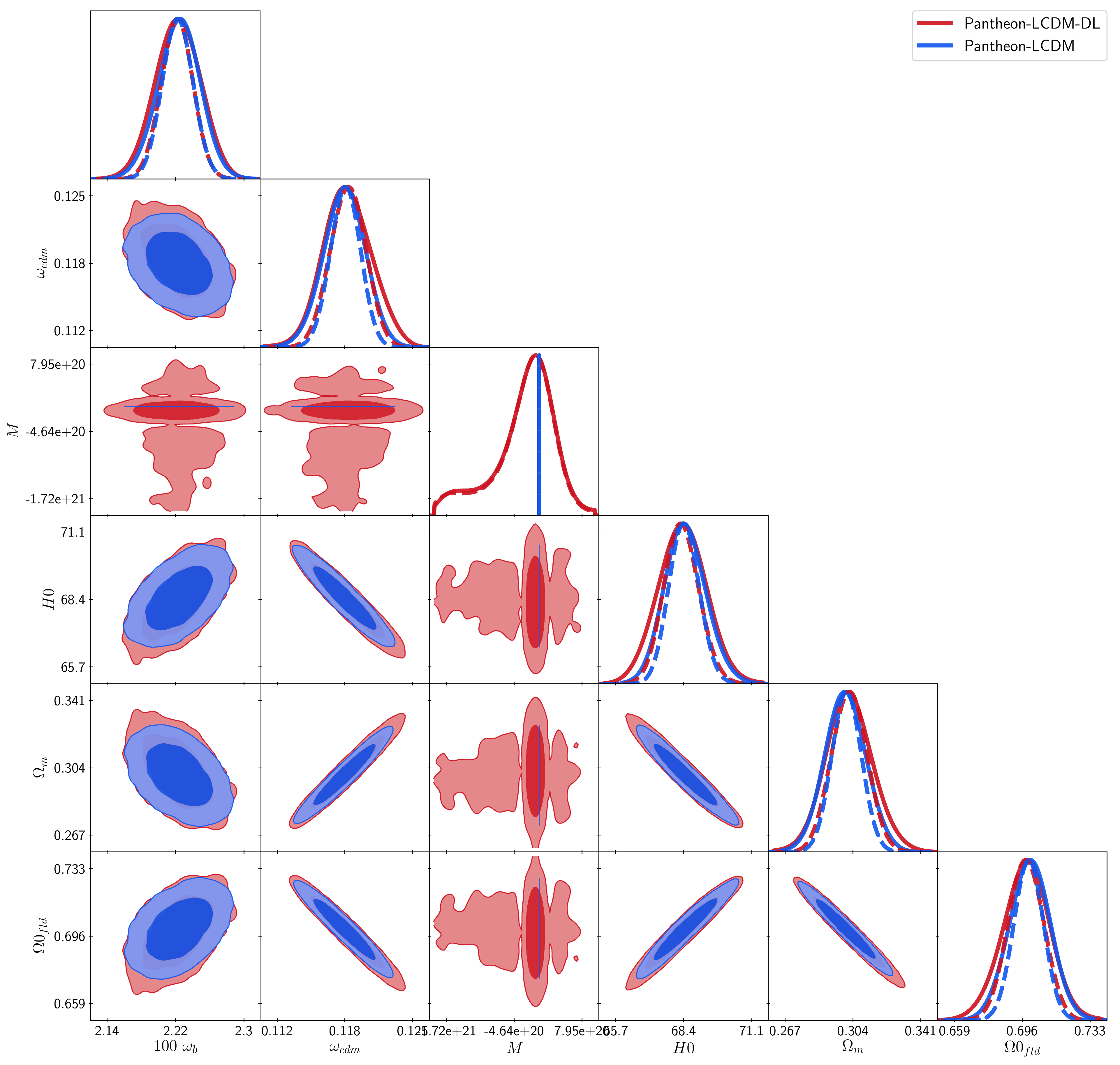}
    \caption{Statistical contours levels for $\Lambda$CDM using observational data (red color) and training deep learning data (blue color).}
    \label{fig:dl_de_lcdm}
\end{figure}

The results of this recipe can be seen in Figure \ref{fig:dl_de_lcdm}.

%%%%%%%%%%%%%%%%%%%%%%%%%%%%%%%%%%%%%%%%%%%%%
%%%%%%%%%%%%%%%%%%%%%%%%%%%%%%%%%%%%%%%%%%%%%

\section{Conclusions}

In this chapter we present how to compute the EoS for dark energy models that lead to an understanding of the problem of the observed cosmic acceleration. Notice that each Bayesian evidence performed will depend on the density data used to develop each cosmological proposal. If we consider more data, the better the statistical analysis will be. Therefore, we expect that future surveys at higher redshift will improve the constraints over the cosmological parameters of the model.

The exploration of these astrophysical surveys have reached a new scenario in regards to the machine learning techniques \cite{Escamilla-Rivera:2019hqt,Munoz:2020gok}. These kinds of techniques allow to explore --without technical problems in the astrophysical devices-- scenarios where the pivot model of cosmology, $\Lambda$CDM, a theoretical framework that accurately describes a large variety of cosmological observables, from the temperature anisotropies of the cosmic microwave background to the spatial distribution of galaxies. 
This scenario has a few free cosmological parameters denoted by fundamental quantities, like the geometry and the Hubble flow, 
the amount and nature of dark energy, and the sum of neutrino masses. If we know the value of these parameters, we will have the capability to improve the fundamental constituents and laws governing our universe. 

Thus, one of the most important goals of modern cosmology is to constrain the value of these parameters with the highest accuracy. Therefore, as an extrapolation between the ideas of the standard cosmostatistics and the use of machine learning techniques will improve even better the constrain of the cosmological parameters without being worried about the intrinsic uncertainties of the data.

%%%%%%%%%%%%%%%%%%%%%%%%%%%%%%%%%%%%%%%%%%%%
%%%%%%%%%%%%%%%%%%%%%%%%%%%%%%%%%%%%%%%%%%%%
\bigskip

\begin{acknowledgments}
CE-R acknowledges the Royal Astronomical Society as FRAS 10147 and the support 
by PAPIIT Project IA100220 and ICN-UNAM projects. 
Also would like to acknowledge networking support by the COST Action CA18108.
\end{acknowledgments}

%%%%%%%%%%%%%%%%%%%%%%%%%%%%%%%%%%%%%%%%%%%%
%%%%%%%%%%%%%%%%%%%%%%%%%%%%%%%%%%%%%%%%%%%%

%%%%%%%%%%%%%%%%%%%%%%%%%%%%%%%%%%%%%%%%%%%%
%%%%%%%%%%%%%%%%%%%%%%%%%%%%%%%%%%%%%%%%%%%%

\end{document}